\documentclass{article}

\usepackage[T1]{fontenc}
\usepackage[utf8]{inputenc}
\usepackage[submission]{ismir} 
\usepackage{amsmath,cite,url}
\usepackage{amssymb}
\usepackage{graphicx}
\usepackage{color}
\usepackage{enumitem}
\usepackage{multirow}
\usepackage{booktabs}
\usepackage{subcaption}
\providecommand{\yw}[1]{\textcolor{black}{{#1}}}
\providecommand{\wyt}[1]{\textcolor{black}{{#1}}}
\providecommand{\minje}[1]{\textcolor{black}{{#1}}}

\title{User-Guided Generative Source Separation}

\multauthor
  {Yutong Wen \hspace{1cm} Minje Kim \hspace{1cm} Paris Smaragdis}
  {University of Illinois at Urbana-Champaign\\
  {\tt\small yutong12@illinois.edu}
  }

\def\authorname{Y. Wen, M. Kim, and P. Smaragdis}

\usepackage[bookmarks=false,pdfauthor={\authorname},pdfsubject={\pdfsubject},hidelinks]{hyperref}

\begin{document}

\maketitle

\begin{abstract}
Music source separation (MSS) aims to extract individual instrument sources from their mixture. 
While most existing methods focus on the widely adopted four-stem separation setup (vocals, bass, drums, and other instruments), this approach lacks the flexibility needed for real-world applications. 
To address this, we propose GuideSep, a diffusion-based MSS model capable of instrument-agnostic separation beyond the four-stem setup.
GuideSep is conditioned on multiple inputs: a waveform mimicry condition, which can be easily provided by humming or playing the target melody, and mel-spectrogram domain masks, which offer additional guidance for separation. 
Unlike prior approaches that relied on fixed class labels or sound queries, our conditioning scheme, coupled with the generative approach, provides greater flexibility and applicability.
Additionally, we design a mask-prediction baseline using the same model architecture to systematically compare predictive and generative approaches. 
Our objective and subjective evaluations demonstrate that GuideSep achieves high-quality separation while enabling more versatile instrument extraction, highlighting the potential of user participation in the diffusion-based generative process for MSS.
Our code and demo page are available at \url{https://yutongwen.github.io/GuideSep/}.
\end{abstract}






\section{Introduction}\label{sec:introduction}

Music source separation (MSS) aims to separate a mixture audio into its constituent sources, typically defined by the instrument. 
Since the 2015 Signal Separation Evaluation Campaign (SiSEC)~\cite{on2015vdbo}, the MSS community has largely focused on supervised models to separate songs into four stems: vocals, bass, drums, and others that includes all remaining instruments, a setup commonly referred to as VBDO. 
Under this framework, numerous recent deep neural network (DNN) models have significantly advanced performance~\cite{rouard2023hybrid, chen2024music, sawata2024whole, tong2024scnet, takahashi2020d3net, luo2023music, lu2024music}.
While this setup provides a convenient benchmark, it lacks the flexibility needed for real-world applications: ideally, MSS systems should be able to extract any target instrument of interest.

In this regard, several works have extended MSS beyond the VBDO setup. 
To enable the separation of arbitrary instruments, the model must first be provided with a condition specifying the target instrument, such as instrument class labels~\cite{meseguer2019conditioned, slizovskaia2019end, seetharaman2019class, samuel2020meta}. In ~\cite{meseguer2019conditioned, seetharaman2019class} this conditioning method is shown to work for the VBDO setup, whereas \cite{slizovskaia2019end} extends this approach to 13 instruments.
However, class labels can be vague, as instruments like the guitar may exhibit significant variability within the same label. Moreover, new instrument classes require re-training.
Another approach, query-based MSS conditions the model using a sound example, where the model extracts sources similar to the example~\cite{smaragdis2009user, lee2019audio, manilow2020hierarchical, watcharasupat2024stem, chen2022zero, wang2022few}. 
For instance, Watcharasupat et al.~\cite{watcharasupat2024stem} designed a lightweight model capable of instrument-agnostic separation using a single query, while Wang et al.~\cite{wang2022few} developed a model that accepts up to five queries to improve performance stability. 
Despite its potential \minje{to provide rich information about the target source, query-based separation} may be limited in real-world applications where high-quality queries are unavailable.
Additionally, MSS models can be conditioned on MIDI score of the target instrument~\cite{ewert2017structured, miron2017monaural, gover2020score, munoz2019online, hung2021transcription}. 
While MIDI information provides a strong and accurate cue, it is often unavailable in many real-world scenarios, such as pop music. 
Bryan et al.~\cite{bryan2014isse, bryan2013interactive, bryan2013efficient, bryan2013refinement} proposed an alternative method where users sketch a rough mask on the spectrogram of the mixture to indicate the target source. 
However, this approach can suffer from ambiguity, as identifying the target instrument's region in the mixture spectrogram is often challenging.
Smaragdis et al.~\cite{smaragdis2009separation} leverages humming as a guidance to separate a target source. Unlike label-based or sound query conditioning, humming offers users greater flexibility when interacting with the system.

In this work, we propose a guided separation (GuideSep) method, a conditional complex-spectrogram domain diffusion model designed to address music source separation beyond the VBDO setup in an instrument-agnostic manner. 
Building on the observations of existing methods for MSS beyond VBDO, we condition the diffusion model on multiple inputs: a waveform \yw{mimicry to a target source} and mel-spectrogram domain masks. 
While MIDI score information is often difficult to obtain in real-world scenarios, users are capable of providing a \yw{mimicry} by humming or playing the target melody with an instrument of their choice. 
Additionally, we introduce a rough mask on the mel-spectrogram for the users to further inform the model of the region to focus on. 
During inference, either or both conditions can be utilized, offering users a flexible way to specify the target source for separation from the mixture.
\yw{Our diffusion model is built on EDMSound~\cite{zhu2023edmsound}, a complex-spectrogram domain diffusion method designed for both unconditional and label-conditioned audio generation. We modify the model backbone to support multiple conditioning inputs.}

\yw{Traditionally, audio source separation has been tackled using predictive models\footnote{Some literature refers to predictive models as discriminative or deterministic. Lemercier et al.~\cite{lemercier2025diffusion} note that predictive models encompass both concepts.}, which map mixture input to an estimated clean output by minimizing a point-wise loss function~\cite{luo2019conv, hu2020dccrn, park2022manner}. 
While predictive models often struggle with residual noise, artifacts~\cite{pirklbauer2023evaluation} in enhancement tasks, generative models have the potential to produce cleaner results by directly or indirectly modeling the clean prior. 
In recent years, significant progress has been made in applying generative models to audio separation tasks, particularly in speech enhancement and separation~\cite{postolache2023latent, subakan2018generative, chen2023sepdiff, lutati2023separate, scheibler2023diffusion, scheibler2024universal, guo2024variance, richter2023speech}. 
While most music source separation (MSS) methods are still predictive, a few generative approaches have begun to emerge.}
For instance, Ge et al. proposed a flow-based model, InstGlow~\cite{zhu2022music}, which leverages the priors of clean sources to improve separation results within the VBDO setup.
Additionally, multi-source diffusion models have been proposed for simultaneous music source separation and generation~\cite{mariani2023multi, karchkhadze2024simultaneous}. 
These approaches employ a multi-channel diffusion process to model the joint distribution of individual sources and condition on the mixture to sample individual sources during inference, enabling separation.
While this formulation provides control over which instrument to synthesize or separate, it is limited to the specific set of instruments the model is trained on.


While there is growing interest in applying generative methods to MSS, to the best of our knowledge, no prior work has systematically compared generative methods with their direct counterparts.
In this work, we address this gap by designing a mask-prediction baseline that shares the exact same model backbone as our diffusion model. 
We then conduct a systematic evaluation to analyze the differences between the two approaches.

Our contributions can be summarized as follows: 1) We propose GuideSep, \yw{one of} the first diffusion-based models designed to address music source separation beyond the VBDO setup \yw{and we release the codebase} 2) We introduce versatile, instrument-agnostic conditions—waveform mimicry conditions and mel-spectrogram domain masks—that are more practical for real-world applications 3) We design a mask-prediction baseline using the same model architecture and conduct a systematic evaluation to analyze the differences between predictive and generative approaches.




\begin{figure}[htbp]
    \centering
    \begin{subfigure}[t]{0.24\textwidth}
        \centering
        \includegraphics[width=\textwidth]{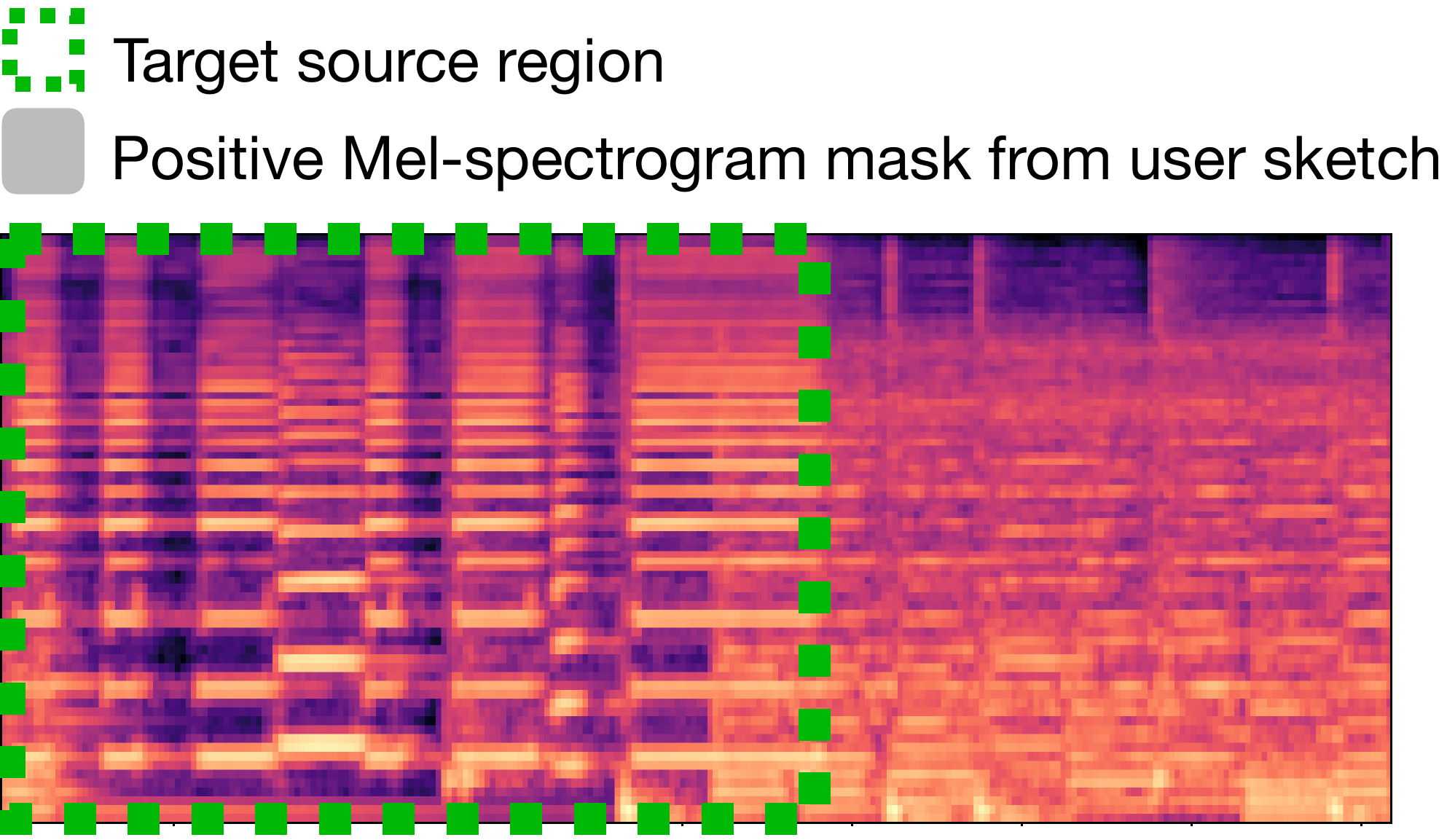}
        \caption{Mel-spectrogram of the mixture}
        \label{fig:sub1}
    \end{subfigure}
    \hfill
    \begin{subfigure}[t]{0.23\textwidth}
        \centering
        \includegraphics[width=\textwidth]{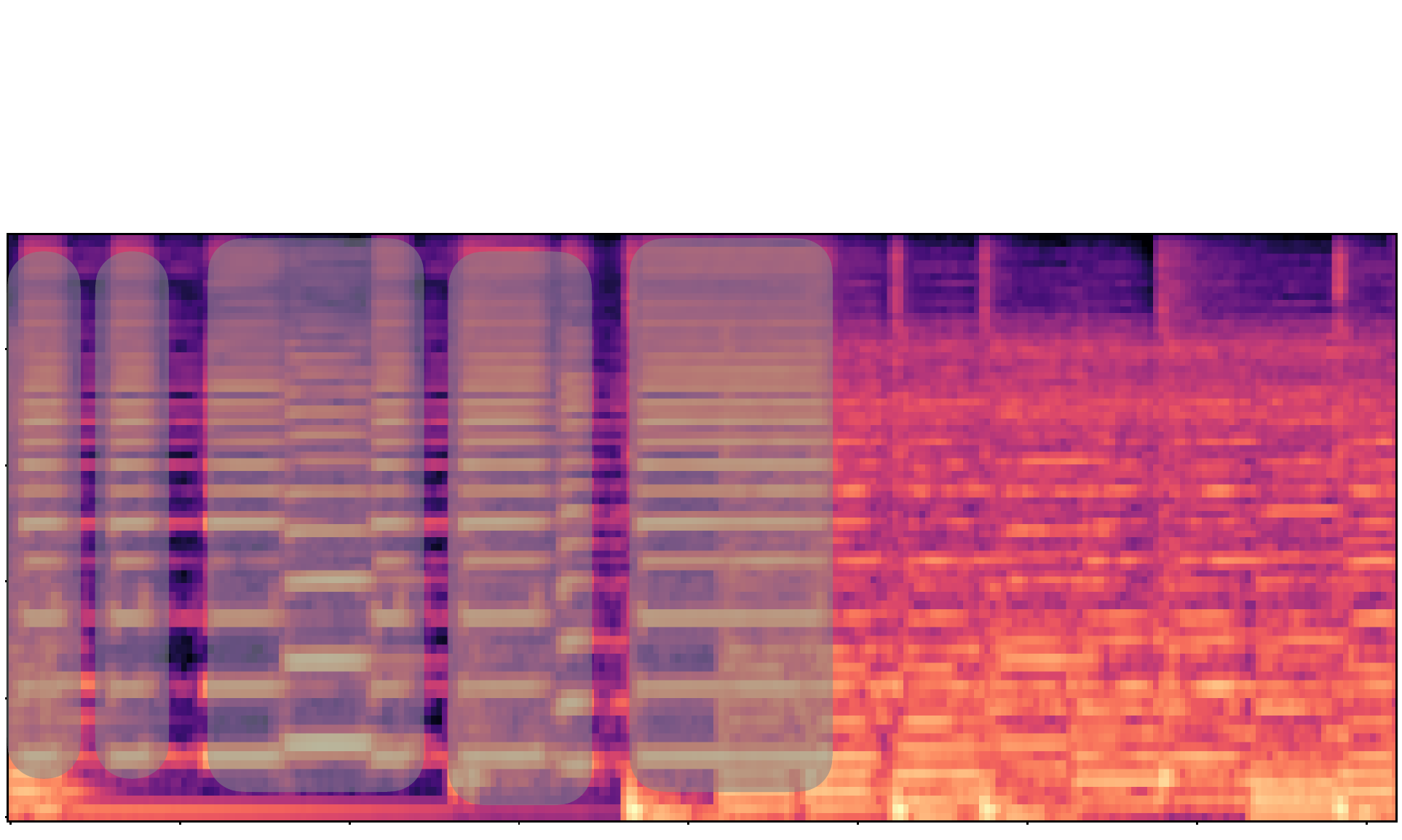}
        \caption{Mel-spectrogram of the mixture with user Mel-mask overlay}
        \label{fig:sub2}
    \end{subfigure}
    \caption{Illustration of an example of positive user-input Mel-spectrogram mask.}
    \label{fig:mel_mask}
\end{figure}

\section{The Proposed GuideSep Method}
\label{sec:method}

GuideSep is a diffusion model conditioned by user input.
Our approach leverages users' input describing a source, i.e., \yw{the raw waveform of user mimicry to a target source} as well as a rough mask in the mel-spectrogram domain. 

\subsection{Condition signals}\label{sec:cond_sig}


\subsubsection{Mimicry condition}

The \yw{mimicry} guidance is a user-provided time-domain waveform, such as a hummed rendition of the target melody or the melody played on another instrument.
Due to the lack of real-world data for training, we simulate the mimicry guidance by converting the ground-truth MIDI score of the target source to audio using the FluidSynth library~\cite{henningsson2011fluidsynth}. 
Real-world \yw{mimicry} inputs often include off-pitch notes, imperfect timing, and limitations in note range. 
\yw{Additionally, since many instruments and vocal mimicry are monophonic, }
it poses significant challenges when extracting polyphonic sources, e.g., guitar or piano.
\yw{We simulate user input via various data augmentation techniques}:
\begin{itemize}[topsep=0pt, itemsep=0pt, parsep=0pt, leftmargin=0.5cm] 
    \item \textbf{Off-pitch} melodies are simulated by introducing perturbations to the MIDI notes. 
    Each note has a $50\%$ probability of being pitch-bent, whose amount is randomly sampled from a uniform distribution ranging from $-0.4$ to $+0.4$ semitones.
    \item \textbf{Imperfect timing} is simulated by introducing variations in the timing of MIDI notes. With a $40\%$ probability, the start and end times of a note are shifted by up to $\pm 30$ milliseconds. The time shift of a note will also be applied to its following notes.
    \item \textbf{Limitation of note range} is simulated by randomly shifting MIDI notes up or down by one octave with a $50\%$ probability.
    \item \textbf{Extraction using non-polyphonic instruments:} We restrict the condition melody to be monophonic to reflect real-world limitations of many instruments \wyt{and humming}. It encourages the model to infer missing notes of the target source using other side information, such as the mel-spectral mask. \yw{The choice of a monophonic condition was driven by our focus on human voice guidance; however, this is a limitation of the training data rather than the algorithm itself.}
\end{itemize}


\subsubsection{Mel-spectral masks}
Our second conditioning input is the user-created mask in the mel-spectrogram domain, that distinguishes regions corresponding to the target source from those of background sources. 
While being conceptually aligned with \cite{bryan2014isse}, GuideSep uses it to condition a deep generative model. Specifically, we define two types of masks: positive and negative masks to indicate the target and background source regions, respectively. 
The mel-spectrogram domain is chosen due to its greater interpretability and easier identification of the sources compared to the Fourier transform's linear frequency scale.

Figure~\ref{fig:mel_mask} illustrates the process of creating a mel-spectrogram mask based on user input. 
\yw{We implement a user interface where} users can sketch on the mel-spectrogram of the mixture with \yw{different brush size and confidence level} to indicate regions they believe correspond to the target source or background music. 
In practice, user-provided masks may exclude portions of the target source or unintentionally include regions of background sound. 
Additionally, in many cases, the target source significantly overlaps the background sources, further complicating the masking process. 
To simulate these real-world imperfections during training, we generate synthetic user input masks by applying a Gaussian filter, whose standard deviation ranges between $4$ to $6$, to the ground-truth mel-spectrograms of the target source and the residual sources.
In addition, we randomly drop out $40\%$ of patches.

\subsection{Conditional complex spectrogram diffusion}
Diffusion probabilistic models (DPMs)~\cite{ho2020denoising, sohl2015deep} consist of two key processes: progressively corrupting training data by adding noise until it approximates a normal distribution, and learning to reverse each step of this noise corruption using the same functional form. 
These models can be generalized as score-based generative models~\cite{song2020score}, which utilize an infinite number of noise scales, enabling both the forward and backward diffusion processes to be described by stochastic differential equations (SDEs). 
During inference, the reverse SDE is employed to generate samples numerically, starting from a standard normal distribution.

\textbf{Complex spectrogram diffusion with EDM: }
Our work is based on EDMSound~\cite{zhu2023edmsound}. 
We train our diffusion model using the EDM framework~\cite{karras2022elucidating}, which reformulates the diffusion SDE in terms of noise scales rather than drift and diffusion coefficients. 
To ensure that the inputs of the neural network are appropriately scaled within the range \([-1, 1]\), as required by the diffusion models, we apply an amplitude transformation to the complex spectrogram inputs. 
Specifically, we use \(\tilde{c} = \beta|c|^{\alpha}e^{i\angle{c}}\), as proposed in~\cite{richter2023speech, breithaupt2010analysis}, where \(\alpha \in (0, 1]\) is a compression factor that emphasizes time-frequency bins with lower energy, \(\angle{c}\) denotes the phase of the original complex spectrogram $c$, and \(\beta \in \mathbf{R}_+\) is a scaling factor that normalizes amplitudes approximately to the range \([0, 1]\).

\begin{figure}[t]
  \centering
  \includegraphics[alt={ISMIR 2025 template example image},width=1\linewidth]{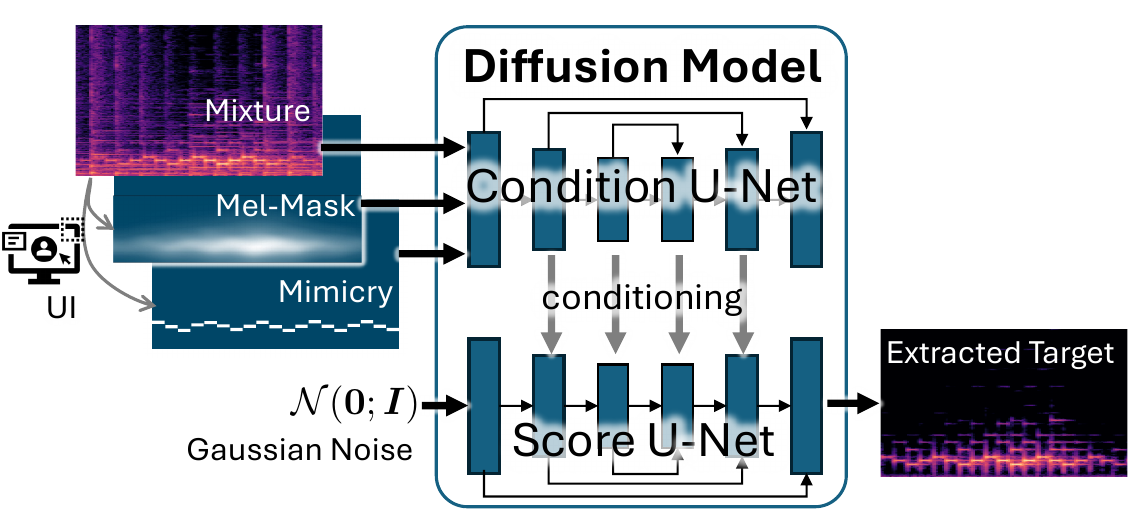}
  \caption{Overview of the GuideSep at inference time. Our model accepts mimicry condition and mel-spectrogram domain masks as guidance \yw{from users} to extract the target source from the mixture.}
  \label{fig:outline}
\end{figure}

\textbf{Adding conditions to EDMSound: }
To adapt EDMSound for target sound extraction, we modify the network to accept conditional inputs, including the mixture signal, \yw{mimicry} signal, and spectral masks. 
Rather than modeling \( p(\mathbf{s}|\mathbf{c}) \), where \( \mathbf{s} \) is the target source and \( \mathbf{c} \) is an instrument  label, we instead model \( p(\mathbf{s}|\mathbf{c}_{\text{mix}}, \mathbf{c}_{\text{mimicry}}, \mathbf{c}_{\text{masks}}) \). 
Here, \( \mathbf{c}_{\text{mix}} \) corresponds to the music mixture represented in the complex-valued short-time Fourier transform (STFT) domain, while \( \mathbf{c}_{\text{mimicry}} \) denotes the \yw{mimicry} condition in the form of magnitude STFT, assuming phase information is a distraction when it comes to representing \yw{spectrum} information. Finally, \( \mathbf{c}_{\text{masks}} \) refers to the normalized magnitudes of mel-spectrogram masks, ranged between $0$ and $1$.

\textbf{The proposed architecture:} Building on insights from prior works~\cite{serra2022universal, scheibler2024universal, wu2024music, garcia2024sketch2sound}, we design our model as depicted in Figure~\ref{fig:outline}. 
The architecture comprises two primary U-Net structures. 
The first one, referred to as the score U-Net, aligns with the original U-Net used in EDMSound. If it were not for conditioning input, this part of the model performs a blind audio synthesis by taking a Gaussian noise sample. The second module, the condition U-Net, is introduced to tame this otherwise entirely generative behavior of the score U-Net. The condition U-Net is dedicated to processing all conditional inputs, including the mixture. 
These two U-Nets are connected so that the output of each layer in the condition U-Net is element-wise added to its corresponding layer in the score U-Net, spanning both the downsampling and upsampling layers.
Since the mel-scale masks are in a different frequency dimension compared to the magnitude and complex spectrograms, we introduce a simple 1-hidden-layer neural network to project the mel-frequency axis onto the spectrogram frequency axis. 
Since there are three conditions in the form of a spectrogram---mixture, mimicry, and projected masks--- we concatenate them along the channel dimension and feed as input to the condition U-Net.
Our U-Net architecture is adapted from Imagen~\cite{saharia2022photorealistic}, chosen for its high sample quality, rapid convergence, and memory efficiency. 

\textbf{Loss function:}
During training, we optimize the model using preconditioned denoising score matching, following~\cite{karras2022elucidating}. 
The training objective is formulated as
\begin{equation*}
\mathbb{E}_{\mathbf{s}} \mathbb{E}_{\mathbf{n}} \left[\lambda(\sigma)\!\left\lVert D\!\left(\mathbf{s}\!+\!\mathbf{n}; \sigma, \mathbf{c}_{\text{mix}}, \mathbf{c}_{\text{mimicry}}, \mathcal{M} (\mathbf{c}_{\text{masks}})\!\right) - \mathbf{s} \right\rVert^2_2\right]\!, 
\end{equation*}
\yw{where $D(\cdot)$ is the EDM weighted neural network,
$\sigma$ is the noise level, $\lambda(\cdot)$ is the loss weighting which is $(\sigma^2 - \sigma^2_{\text{data}})/(\sigma \cdot \sigma^2_{\text{data}})$ for the EDM framework,} 
$\mathcal{M}(\cdot)$ denotes the 1-hidden-layer projection network, and \(\mathbf{n} \sim \mathcal{N}(\mathbf{0}, \sigma^2\mathbf{I})\) is Gaussian noise. 

\textbf{Inference:} \yw{Within the EDM framework, the probability flow \minje{ordinary differential equation (ODE)} can be simplified into a nonlinear ODE, allowing the direct use of standard off-the-shelf ODE solvers, such as  high-order Exponential Integrator (EI)-based ODE solvers~\cite{lu2022dpm}, specifically multistep DPM solvers~\cite{lu2022dpm, lu2022dpmpp}, for sampling as in EDMSound.}

\begin{table*}[h!]
\centering
\renewcommand{\arraystretch}{1.1}
\resizebox{\textwidth}{!}{
\begin{tabular}{cccccccccccc}
\toprule
Model & Piano & Guitar & Bass & Strings & Brass & Synth & Pipe & Reed & Organ & \begin{tabular}[c]{@{}c@{}}Chromatic \\ Percussion\end{tabular} & Overall \\ \hline
Ours (full)                  & \bf{8.34$\pm 0.11$} & \bf{10.53$\pm 0.09$} & \bf{11.97$\pm 0.12$} & \bf{9.64$\pm 0.12$} & \bf{9.15$\pm 0.38$} & \bf{9.25$\pm 0.20$} & \bf{15.58$\pm 0.27$} & \bf{13.78$\pm 0.24$} & \bf{13.44$\pm 0.22$} & \bf{11.53$\pm 0.36$} & \bf{10.46} \\
Baseline (full)              & 7.03$\pm 0.09$ & 8.72$\pm 0.08$  & 8.69$\pm 0.06$  & 9.06$\pm 0.13$ & 8.03$\pm 0.31$ & 8.00$\pm 0.17$ & 14.99$\pm 0.26$ & 11.94$\pm 0.23$ & 11.25$\pm 0.23$ & 9.62$\pm 0.31$  & 8.74  \\ \hline
Ours (mimicry only) & 7.46$\pm 0.11$ & 9.96$\pm 0.10$  & 11.19$\pm 0.13$ & 8.63$\pm 0.14$ & 7.95$\pm 0.45$ & 8.13$\pm 0.23$ & 14.43$\pm 0.34$ & 13.14$\pm 0.27$ & 12.39$\pm 0.25$ & 8.74$\pm 0.43$  & 9.60 \\
w/ pseudo-masks              & 7.99$\pm 0.11$ & 10.18$\pm 0.10$ & 9.87$\pm 0.15$  & 8.72$\pm 0.15$ & 8.21$\pm 0.40$ & 8.19$\pm 0.25$ & 14.81$\pm 0.31$ & 13.20$\pm 0.26$ & 12.20$\pm 0.29$ & 8.26$\pm 0.55$  & 9.56 \\
Ours (positive mask only)    & 7.86$\pm 0.11$ & 10.17$\pm 0.09$ & 11.45$\pm 0.13$ & 9.48$\pm 0.12$ & 8.97$\pm 0.38$ & 9.14$\pm 0.19$ & 15.19$\pm 0.28$ & 13.42$\pm 0.25$ & 13.08$\pm 0.23$ & 11.09$\pm 0.38$ & 10.09 \\ 
Ours (humming)*              & -    & -     & -     & -    & -    & -    & -     & -     & -     & -     & 13.61 \\ \hline
Frequency (\%)         & 20.71   & 27.89   & 17.79     & 15.23   & 2.65    & 4.74    & 2.72    & 3.06     & 3.43    & 1.78    & - \\ \hline
\bottomrule
\end{tabular}}
\caption{SDR (dB) results with 95\% confidence interval (higher values indicate better performance) for ten instrument classes in the Slakh2100 test split. The results include GuideSep (our method) under various input conditions and the mask-prediction baseline. The best scores are highlighted in bold. For asterisk (*) please refer to Section~\ref{sec:ablation}.}
\label{tab:exp}
\end{table*}

\section{Experiment}

We conduct experiments using the Slakh2100 dataset~\cite{manilow2019cutting} augmented by MoisesDB ~\cite{pereira2023moisesdb} for training. The Slakh2100 dataset provides an official train-validation-test split, which we utilize as well. 
\yw{We evaluate our model's performance using the widely adopted signal-to-distortion ratio (SDR) metrics~\cite{vincent2006performance, scheibler2022sdr}.}

\subsection{Training and model details}\label{sec:training}

\subsubsection{The datasets}
Slakh2100 is a synthetic dataset of waveform-MIDI-aligned music dataset containing 2,100 tracks in total around 145 hours of audio. 
In our training process, instead of using the original mix from the dataset, we generate training data through random mixing. 
This way allows for nearly infinite variations of training samples. 
While this approach may result in the loss of some musical context, previous work~\cite{jeon2024does} has demonstrated that random mixing can improve MSS model performance.
To enhance our model's performance on real-world music, we utilize the MoisesDB dataset~\cite{pereira2023moisesdb} to construct background sources. 
MoisesDB is a comprehensive multitrack dataset designed for source separation beyond 4-stems, featuring 240 previously unreleased songs by 47 artists across twelve high-level genres, in total approximately 14 hours of audio. 
During random mixing, we randomly select 3 to 6 sources from the MoisesDB dataset to serve as background music, while the target source is drawn from the Slakh2100 dataset. 
The background and target sources are mixed at signal-to-noise ratios (SNR) ranging from \(-5\) dB to \(5\) dB. All The input audio is converted to single channel and resampled to 16 kHz, and then trimmed or padded to around $4.1$ seconds for batched training.

\subsubsection{Dropout strategies}
To ensure that the model can process any combination of input types, we incorporate dropout strategies during training.
This allows the model to operate with an incomplete set of conditions, such as the mimicry-only or mel-spectrogram-mask-only cases. To this end, 
we randomly drop out either the mimicry condition or mel-spectrogram masks, ensuring that the model learns to predict the target source even when provided with partial conditioning information.
Additionally, we empirically observe that the model benefits from a \yw{mimicry}-only conditioned synthesis tasks, which happens when we randomly drop the mixture input $\mathbf{c}_\text{mix}$ during training. 
This encourages the model to infer the target source from melodic guidance alone.
Specifically, during training, we drop \(30\%\) of the mimicry condition, \(70\%\) of the mel-masks, and \(10\%\) of the mixture. 
The high dropout rate for mel masks is intentional \yw{and tuned using the validation split}, as they provide a strong cue to the target source. 
By reducing their presence, the model is encouraged to focus more on learning from the \yw{mimicry} condition.

\subsubsection{The model architecture}
For both score and condition U-Net modules, we utilize an efficient U-Net architecture adapted from the open-source Imagen implementation\footnote{https://github.com/lucidrains/imagen-pytorch}, which is known for memory efficiency and fast convergence. Both U-Nets incorporate downsampling and upsampling blocks, each containing two ResNet blocks with a self-attention layer that uses two attention heads. The bottleneck dimension is 128. The complete model has 93.3 million trainable parameters.

The input to the condition U-Net consists of three types: a complex spectrogram $\mathbf{c}_\text{mix}$, a magnitude spectrogram $\mathbf{c}_\text{mimicry}$, both with a window size of 512 samples and a hop size of 256 samples, and the mel-spectrogram masks $\mathbf{c}_\text{mask}$, which share the same hop size and consist of 80 mel-frequency bins. Eventually, it is a five-channel spectrogram input: two for the complex spectrogram, one for the magnitude spectrogram, and two for the positive (i.e., target source) and negative (i.e., background music) masks. 

The score U-Net, as an autoencoder, defines a two-channel spectrograms as its input and output representation, where the two channels represent the real and imaginary components of the complex spectrogram, respectively. Note that in the very beginning of the sampling process, the input spectrogram to the score U-Net is noise sampled from Gaussian. 
Additionally, we condition the network on logarithmically scheduled noise levels \(\sigma\).

\subsubsection{Inference}
For inference, we employ an EI-based DPM sampler~\cite{lu2022dpm,lu2022dpmpp}. 
To ensure compatibility between the EDM framework samplers and arbitrary training objectives during inference, we implement input rescaling as needed. 
Specifically, we rescale both the noisy inputs and noise levels to align with the network's original training-time scales. 
The results, presented in Section~\ref{sec:result}, are obtained using an 8-step sampler configuration.

\subsubsection{Training details}
Our model is trained with a batch size of 36 and a learning rate of $1\times 10^{-4}$ using the Adam optimizer. 
The training process runs for 300k updates.
We used two NVIDIA L40S GPUs, and trained for ten days.

\subsection{Baselines}
\yw{To the best of our knowledge, no existing work offers a fair comparison, as our method introduces a novel conditioning approach.}
However, we design a traditional mask-prediction model to compare the proposed generative approach against. 
The baseline shares the same twin U-Net architecture and structural details as our diffusion backbone. 
In particular, the input to the score U-Net portion is the magnitude spectrogram of the mixture, while the input to the condition U-Net consists of the magnitude spectrogram of the mimicry condition and the masks. 
The model outputs a non-binary mask by applying a sigmoid function after the output layer, which is then used to compute the target source magnitude spectrogram through element-wise multiplication with the input mixture magnitude spectrogram.
The final waveform is reconstructed by combining the predicted magnitude spectrogram with the phase information from the original mixture.
We train the mask-prediction baseline using the L2 reconstruction loss in the \yw{magnitude spectrogram domain}, with the same learning rate, batch size, and number of updates as our diffusion model.
Note that, due to the absence of time-step conditional inputs and differences in input channels, the mask-prediction baseline contains 80.3 million parameters.


\section{Evaluation and Discussion}\label{sec:result}

We evaluate our model on the official test split of the Slakh2100 dataset. 
The \yw{mimicry} condition signals are synthesized as described in Section~\ref{sec:cond_sig}, using randomly selected virtual instruments from the FluidSynth library.
Similarly, the positive and negative masks are simulated following the same procedure outlined in Section~\ref{sec:cond_sig}. 
For evaluation, we group the instrument classes in Slakh2100 into ten broader categories, where drum tracks are excluded from the target sources, because our \yw{synthesis method} does not apply to them. The evaluation results are presented in Table~\ref{tab:exp}.

In the first two rows in Table~\ref{tab:exp}, we present the results of our model and the mask-prediction baseline. 
Both models utilize the \yw{mimicry} condition and mel-spectrogram masks during inference, denoted with `(full)' in the table. 
The results demonstrate that our model consistently outperforms the mask-prediction baseline across all instrument classes. 
Given that the mask-prediction baseline shares the same model backbone, training data, and configuration as our diffusion model, the performance gap highlights the benefits of using diffusion approach. 
In the listening test, we observe that the mask-prediction baseline often \minje{reconstructs target sources which still contain interferences}. 
In contrast, while our diffusion model may occasionally exhibit inexact timbre, it generally generates cleaner target sources. 
This can be attributed to the diffusion model learning a prior distribution of clean sources, which biases its outputs toward cleaner results. \minje{Although our findings align to the well-known behavior of generative models, our experiments are limited to the particular choice of the diffusion model and a masking-based baseline with a matching architecture, leaving more general arguments to future work.}
We also observe that \minje{both} models work better for the monophonic sources than the polyphonic ones, such as piano, guitar, strings, and synth, where our strictly monophonic mimicry condition is not informative enough. 
As a result, the models may struggle with missing notes from chords, extracting the wrong target instrument, or even extracting multiple instruments when they share a similar melody, which is common in music.
For results on the real-world conditions, please refer to our demo page.

\subsection{Subjective Listening Tests}
In addition to the SDR results, we conduct a subjective listening test to further evaluate our model. 
\minje{We modify webMUSHRA~\cite{schoeffler2018webmushra}} so the test comprises two sections: the first assesses the overall quality of the model's separation results, while the second focuses specifically on evaluating the timbre of the reconstructed target source.
In the first section, each question presents the music mixture as a reference.  
Participants are asked to compare and rate four stimuli: the ground truth, the mixture itself (i.e., the hidden reference), and the predictions from our model and the mask-prediction baseline.  
Participants are unaware that one of the stimuli is the actual ground truth and are instead told that the three stimuli are potential reconstructions of a target source.  
The participants are asked to first identify the mixture and assign it a score of 0, then rate the remaining stimuli (e.g., with 100 being a perfect match) based on how closely they resemble the target source in the mixture.
This part consists of ten trials, with each mixture sample randomly selected from a different instrument class in the Slakh2100 test split. We use the mixture as a reference instead of the ground-truth source in order to measure the listener's opinion on the ``synthesized" source without introducing any prejudice.

To make up the modification introduced in the first part, the second part is dedicated to evaluating the potential artifact specific to the generative models, i.e., the timbre change. This time, each trial presents the ground-truth target source as the reference. Participants compare and rate three stimuli: the hidden reference (i.e., the ground truth itself) and the predictions from our model and the mask-prediction baseline. 
However, ratings are based on timbre similarity to the reference, with 100 indicating an exact match and 0 representing a completely different timbre. 
Participants are instructed to focus solely on the timbre of the target source while disregarding any interference or artifacts.
Both parts use the same set of music samples, but the second part is presented only after participants complete the first part to avoid bias, ensuring they remain unaware that the ground truth was included in the first part.


\begin{figure}[htbp]
    \centering
    \begin{subfigure}[t]{0.259\textwidth}
        \centering
        \includegraphics[width=\textwidth]{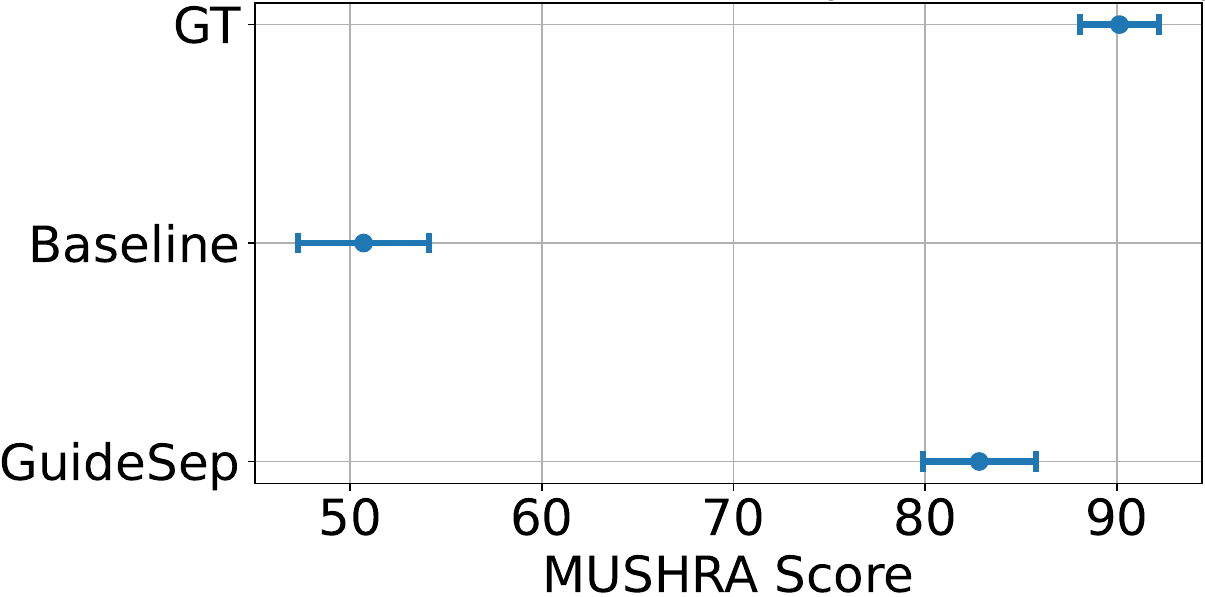}
        \caption{Sec. 1, MUSHRA result on separation quality}
        \label{fig:sub1}
    \end{subfigure}
    \hfill 
    \begin{subfigure}[t]{0.211\textwidth}
        \centering
        \includegraphics[width=\textwidth]{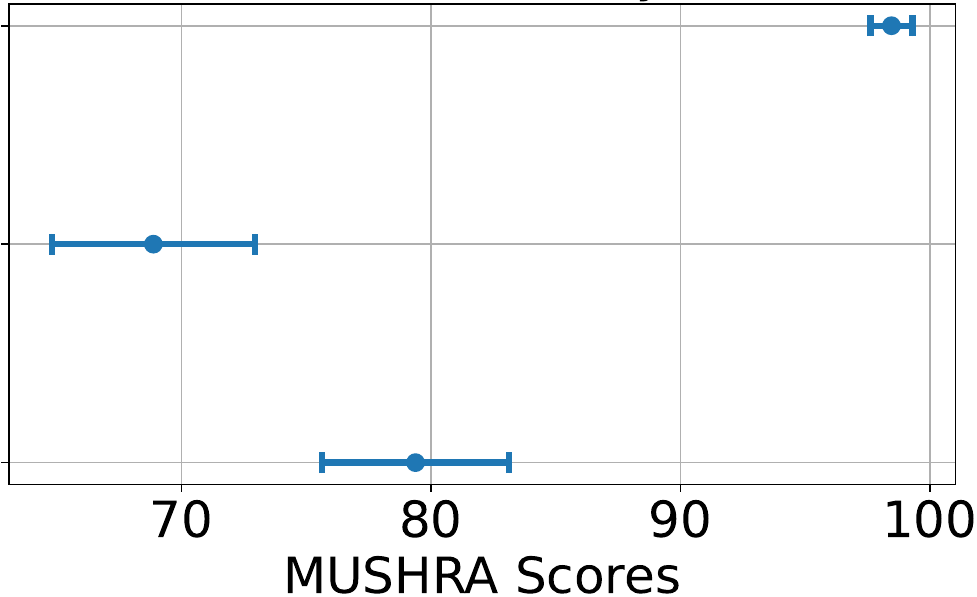}
        \caption{Sec. 2, MUSHRA result on timbre similarity}
        \label{fig:sub2}
    \end{subfigure}
    \caption{Mean MUSHRA Score with 95\% confidence interval of the subjective listening test on separation quality and separation timbre quality.}
    \label{fig:mushra}
\end{figure}

A total of \yw{13} participants took part in the subjective listening test, and the results from both parts are presented in Figure~\ref{fig:mushra}.  
In the first part, where participants rated the separation quality, our model scored $82.82\pm2.95$, the ground truth scored $90.13\pm2.06$, and the mask-prediction baseline scored $50.69\pm3.41$.  
Notably, despite the mask-prediction baseline having a relatively small SDR difference from our model, the listening test revealed a significant gap in perceptual evaluation. 
This suggests that users may perceive a cleaner target source prediction as more satisfactory, even if a slightly noisier prediction achieves a decent sample-wise similarity to the target source.

In the second part, where participants rated timbre preservation, our model scored $79.38\pm3.76$, surpassing the mask-prediction baseline, which scored $68.88\pm4.07$.  
In theory, the mask-prediction baseline could preserve the original timbre better, but our subjective listening test results suggest otherwise. 
Based on the listeners' feedback, we speculate that this outcome is influenced by the nature of the target source extraction task, where multiple sources in a musical piece may share similar melodic patterns.  
As a result, the mask-prediction baseline's output can be contaminated by interfering similar melodies, which can be perceived as a timbral change rather than artifacts. 

\subsection{Ablation}\label{sec:ablation}

Beyond evaluating our model with both conditioning signals, we conduct an ablation study to assess its performance under different input conditions. 

\textbf{Mimicry-only}: We evaluate the model using the `(mimicry only)' setup  (Table~\ref{tab:exp}). 
We observe a slight overall decrease in performance, indicating that while mel-masks contribute to improved performance, the model remains effective even when conditioned solely on the melody signal.

\textbf{Pseudo-masks}: When only the mimicry condition is available, we can generate pseudo mel-masks using the mimicry condition and the mixture. 
Specifically, we use the Gaussian-blurred mel-spectrogram of the mimicry condition as the positive mel-mask and the blurred mixture as the negative mel-mask with the standard deviation set to be 5. 
In Table~\ref{tab:exp}, although the overall SDR score is slightly lower compared to using only the mimicry condition, the model performs better with pseudo-masks for 7 out of 10 instrument classes. 
This suggests that pseudo-masks can generally enhance the model's performance at no additional cost.
The bass class is an exception, likely due to its limited high-frequency content, which sets it apart from other instruments. 
Consequently, the mel-spectrogram mask may be misleading in this case.
A different standard deviation for the Gaussian filter could work better, while it involves an additional hyperparameter search. 

\textbf{Mel-masks-only}: Another case is when only the mel-masks are used for conditioning. 
We observe that the results are generally better than those obtained using only the mimicry condition, indicating that mel-masks serve as highly effective conditioning signals.

\textbf{Humming-only:}
Although in our training, mimicry condition do not include humming, we evaluate our model to assess its generalization to unseen mimicry condition, such as humming. 
\yw{Since we cannot easily synthesize humming from MIDI}, we utilize the HumTrans dataset~\cite{liu2024humtrans}, a MIDI-humming aligned dataset, resulting in an evaluation dataset of approximately \(16.6\) hours. 
Since HumTrans melodies do not coincide with our test songs, an ideal source separation setup is impossible to design. 
Instead, we synthesize background sources by randomly mixing 3 to 6 sources from the MoisesDB dataset, following our training procedure procedure described in Section~\ref{sec:training}. 
The target source is synthesized from the MIDI information aligned to the humming, using the method outlined in Section~\ref{sec:cond_sig} with augmentation. 
As the virtual instruments are sampled from the FluidSynth library, which is not directly comparable to the Slakh2100 benchmark, we report only an overall SDR result, whose mean is \(13.61\) dB. 
This score exceeds the overall SDR result of our model on the Slakh2100 benchmark, demonstrating that the mimicry condition can generalize to humming during inference. 
However, the strong performance could also be attributed to the random mixing used during evaluation, which simplifies the task of target source extraction for the model.

\section{Conclusion}
We introduced GuideSep, a diffusion-based music source separation model that enables flexible, instrument-agnostic separation using waveform mimicry conditions and mel-spectrogram masks, and released the codebase. 
Our results demonstrate that this approach achieves high-quality separation while offering greater adaptability compared to traditional class-based methods. 
Additionally, our comparison with a mask-prediction baseline provides insights into the strengths of generative models for MSS. 
This work highlights the potential of diffusion models in advancing more versatile and user-controllable source separation.

\bibliography{reference}

\end{document}